\documentclass[mypaper,7pt,twoside]{CoAst}
\usepackage{epsf,graphicx,fancyhdr}
\renewcommand{\chaptermark}[1]{\markboth{#1}{}}

\newcommand{\kopf}{\small\itshape Comm. in Asteroseismology \\ Contribution to the Proceedings of the Wroclaw HELAS Workshop, 2008}

\newcommand{\Authors}[1]{\begin{center}\normalsize\bf\sf #1 \end{center}}

\renewcommand{\author}[1]{\begin{center}\normalsize\bf\sf #1 \end{center}}
\newcommand{\Address}[1]{\begin{center}\small\sf #1 \end{center}}

\newcommand{\Session}[1]{{\vspace{3mm}\small \noindent  \hspace*{3mm} Session: } #1 \normalsize}

\newcommand{\Objects}[1]{{\vspace{0mm}\small \noindent  \hspace*{3mm} Individual Objects: } \small #1 \normalsize}

	\newcommand{\threeE}{\small STARS - oscillatins and stellar models \newline}

\newcommand{\discussion}[2]{{\textbf{\small#1:}}{ \small #2 \newline \normalsize}}

\renewenvironment{abstract}{\section*{Abstract}\normalsize\sf}{}
\newcommand{\References}[1]{\begin{flushleft}{\large References\\}\vspace*{2mm}\small #1 \end{flushleft}}

\newcommand{\chapterCoAst}[2]{\chapter[\sf\normalsize #1\\ \footnotesize \hspace*{5mm}by #2 \sf\normalsize][]{#1\\}\rhead[\fancyplain{}{\sf\footnotesize \center{#1}}]{\fancyplain{}{\sffamily\thepage}}\lhead[\fancyplain{\kopf}{\sffamily\thepage}]{\fancyplain{\kopf}{\sf\footnotesize \center{#2}}}}

\newcommand{\figureDSSN}[5]{\begin{figure}[#4]
\centering
\includegraphics*[#5]{#1}
\caption{#2}
\label{#3}
\end{figure}}

\renewcommand{\chaptername}{}

\newcommand{\acknowledgments}[1]{\vspace*{5mm}\noindent  \textbf{Acknowledgments.} #1}

\def\rfr{\smallskip\par\noindent
        \hangindent=7truemm
        \hangafter=1}

\begin{document}
\sf

\chapterCoAst{Examples of seismic modelling}
{A.\,A.\,Pamyatnykh} 
\Authors{A.\,A.\,Pamyatnykh$^{1,2,3}$}
\Address{
$^1$ N.\,Copernicus Astronomical Center, ul.\,Bartycka 18, 00-716 Warsaw, Poland\\
$^2$ Institut f\"ur Astronomie, T\"urkenschanzstrasse 17, A-1180 Vienna, Austria\\
$^3$ Institute of Astronomy, Pyatnitskaya Str.\,48, 109017 Moscow, Russia }

\noindent
\begin{abstract}

Findings of a few recent asteroseismic studies of the main-sequence
pulsating stars, as performed in Wojciech Dziembowski's group in
Warsaw and in Michel Breger's group in Vienna, 
are briefly presented and discussed. 
The selected objects are three hybrid pulsators $\nu$\,Eridani, 
12\,Lacertae and $\gamma$\,Pegasi, which show both $\beta$\,Cephei and
SPB type modes, and the $\delta$\,Scuti type star 44\,Tauri.

\end{abstract}

\Session{ \threeE }
\Objects{$\nu$\,Eri, 12\,Lac, $\gamma$\,Peg, 44\,Tau}

\section*{Introduction}

Main-sequence stars (including the Sun) seem to be ideal 
objects for the seismic modelling, i.e. for construction
- or rather refinement - of the models of individual stars using
data on their multiperiodic oscillations. The structure of the MS
stars has been understood qualitatively quite well but some details
of the structure and some physical processes still need to be
explained. These problems have been specified and discussed in a
very compact and informative form by Marc-Antoine Dupret in his
introductory talk (these Proceedings). By fitting stellar models to
observational data on oscillations we can obtain detailed
constraints on stellar parameters and on physical processes in
interiors.

Most of the results, presented in this contribution, have been
discussed more thoroughly by Dziembowski \& Pamyatnykh (2008) and by
Lenz et al.\,(2008). However, newest observational data (in
particular, for 44 Tau) may change some conclusions of those papers.

\section*{Hybrid stars $\nu$\,Eri, 12\,Lac and $\gamma$\,Peg}
Hybrid stars are main-sequence pulsating variables which show two
different types of oscillations: (i) low-order acoustic and gravity
modes of $\beta$\,Cephei type with periods of about $3-6$ hours, and
(ii) high-order gravity modes of the SPB type with periods of about
$1.5-3$ days. Theoretically, such a behaviour
was predicted for $\ell > 5$ by Dziembowski and Pamyatnykh (1993,
see Figs.\,5 and 6 there) and later, when using newer opacity data,
\figureDSSN{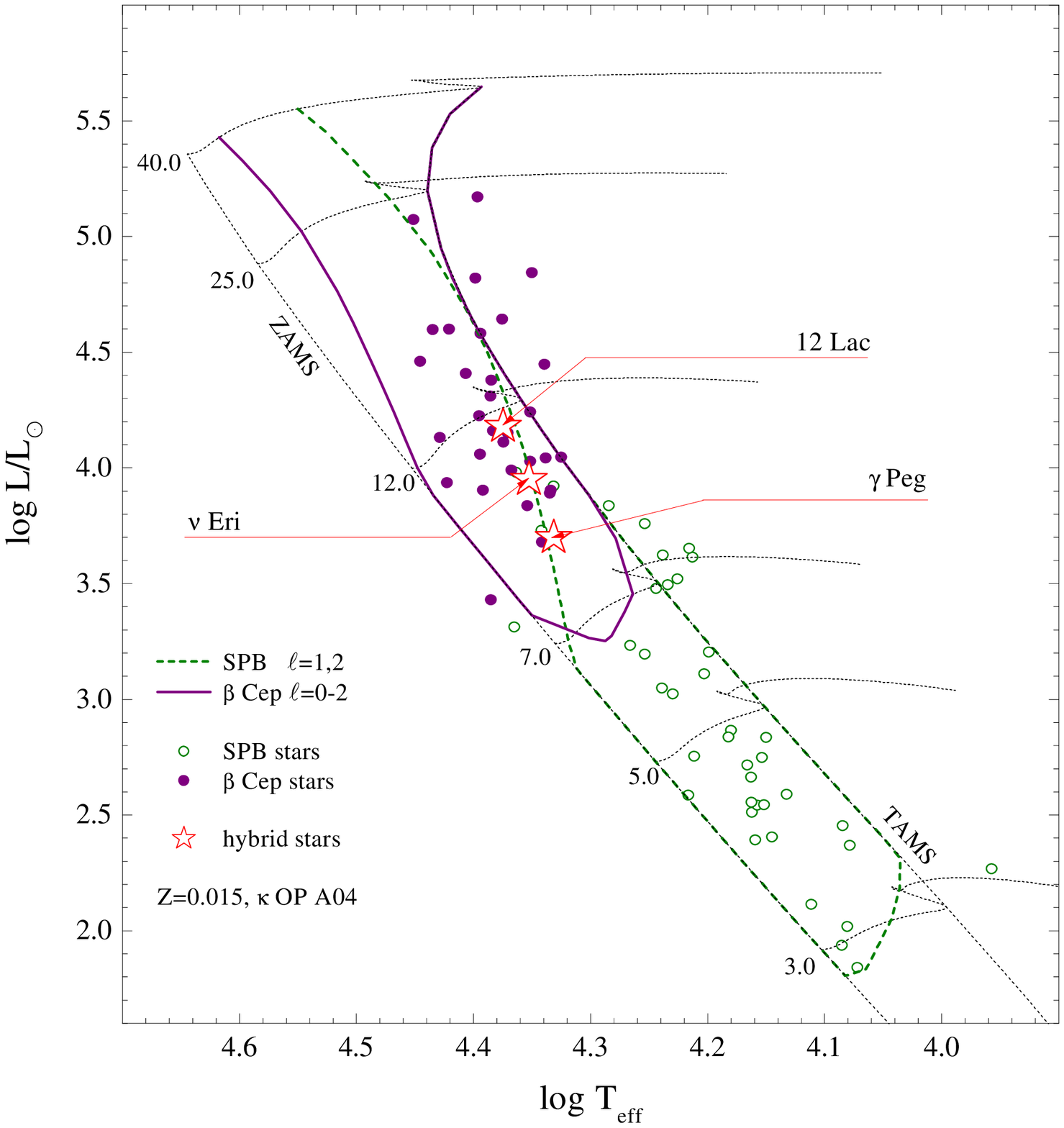}
{
Pulsational instability domains in the upper part of the main
sequence. In the overlapped region of the $\beta$\,Cep type and
SPB type oscillations the hybrid pulsations are expected. OP opacity
data for $Z=0.015$ and for new solar proportions in the heavy
element abundances were used (mixture A04, Asplund et al. 2005). The
positions of observed stars were obtained using catalogs of Stankov
\& Handler (2005, $\beta$\,Cep) and De Cat (2007, SPB). Three
hybrid stars are marked explicitly.
}
{HRD_BCep_SPB_domains}{!h}{clip,angle=0,width=80mm}
also for the low-degree oscillations (Pamyatnykh 1999). In the
HR diagram, the overlapped region of the $\beta\,$Cep and SPB type
pulsations is very sensitive to the opacity data
(see Figs.\,3 and 4 in Pamyatnykh 1999), therefore the modelling
of hybrid star pulsations will allow to test the opacity
of the stellar matter - for example, to choose between two
independent sets of the OPAL and OP data
(Iglesias \& Rogers 1996 and Seaton 2005, respectively). Theoretical
$\beta\,$Cep and SPB instability domains and position of three
confirmed hybrid stars are shown in Fig.\,1. All
three stars are very close to the region where hybrid pulsations are
expected when using the OP opacities.
\figureDSSN{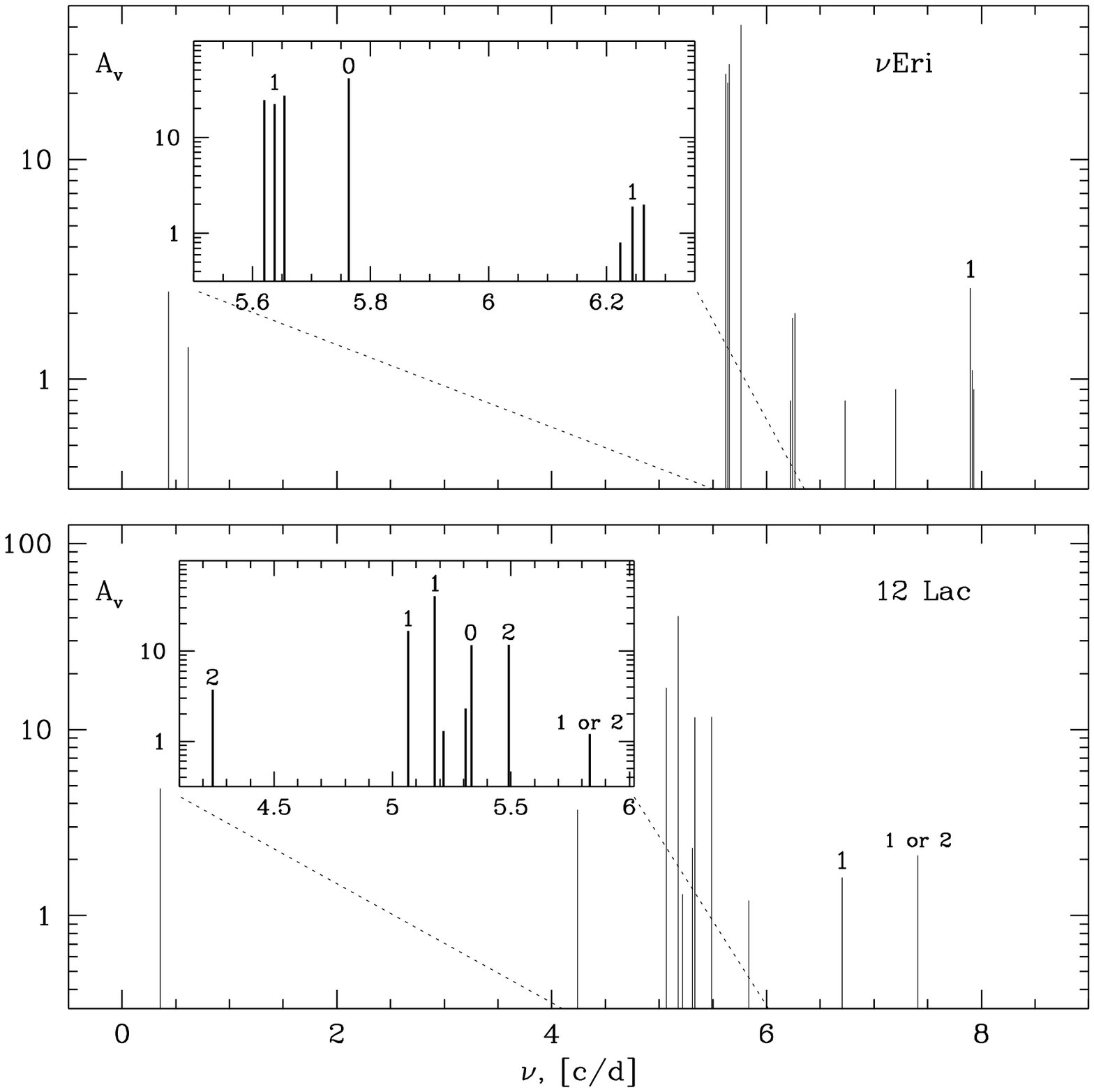}
{
Observational frequency spectra of
{\bf $\nu$\,Eri} and {\bf 12\,Lac} based on Jerzykiewicz et al. (2005) and
Handler et al. (2006) data, respectively. The numbers above
frequency bars mark mode degree values, $\ell$, as inferred from
multicolour photometry (according to Dziembowski \& Pamyatnykh
2008).
}
{NuEri_12Lac_ObsFreq}{!ht}{clip,angle=0,width=77mm}

Fig.\,2 shows the observational frequency spectra of {\bf
$\nu$\,Eri} and {\bf 12\,Lac}. High-frequency modes of the
$\beta\,$Cep type are well resolved, and their degree, $\ell$, and
in some cases also azimuthal order, $m$, are determined from
photometry and spectroscopy. For very slow rotating star $\nu$\,Eri,
one radial mode, two $\ell=1$ triplets (modes ${\rm g}_1$ and ${\rm p}_1$) and
one more mode ($\ell=1, {\rm p}_2$, $\nu=7.89$~c/d) are identified.
For 12\,Lac, one radial mode, two components of a $\ell=1$ triplet
and 5 other $\ell=1$ and/or $\ell=2$ modes are identified.


The seismic models of {\bf $\nu$\,Eri} which fit radial and two
$\ell=1, m=0$ frequencies were constructed for different opacities
(OP and OPAL) and different assumptions about the efficiency of the
overshooting from the stellar convective core. Ausseloos et
al.\,(2004) constructed seismic models of this star which
fit one more mode, $\ell=1, {\rm p}_2$ at $\nu=7.89$~c/d, but it was
necessary to assume rather unrealistic chemical composition
parameters ($X$ and/or $Z$) and a quite effective overshooting.
Moreover, they did not consider the low-frequency modes of the SPB
type pulsations. We argued that the overshooting from the convective core
is ineffective in $\nu$\,Eri (seismic models with overshooting are located outside
the observational error box in the HR diagram) but this result critically depends
on the $T_{\rm eff}$ determination.
\figureDSSN{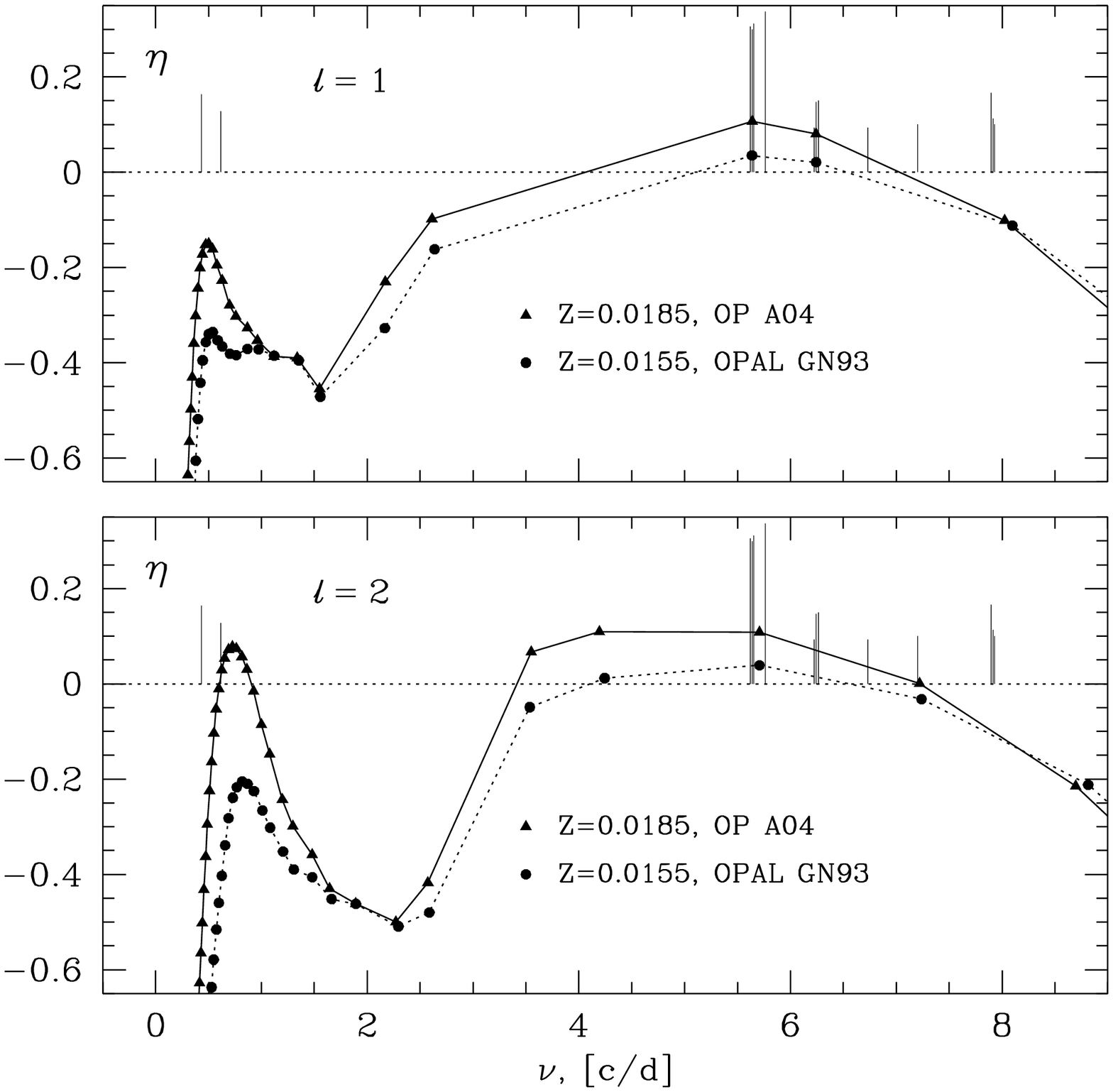}
{
The normalized growth rate, $\eta$,
as a function of mode frequency for {\bf $\nu$\,Eri} seismic model
constructed both with OPAL and OP opacities ($\eta\,>\,0$ for
unstable modes). Both models fit the observational error box ($\log\,L,
\log\,T_{\rm eff}$) in the HR diagram. The frequencies of radial
fundamental and two dipole modes ($\ell=1$, modes ${\rm g}_{1}$ and ${\rm p}_{1}$)
fit the observed values at $5.763$, $5.637$ and $6.244$ c/d,
respectively. The observed frequencies are shown by vertical
lines, with their lengths proportional to the mode amplitudes in a
logarithmic scale. Note that in the OP case there are unstable
low-frequency quadrupole ($\ell=2$) modes, whereas in the OPAL
case all low-frequency modes are stable.
}
{Eta_NuEri_OP_OPAL}{!ht}{clip,angle=0,width=77mm}

Fig.\,3 illustrates the effect of the opacity choice between the
OPAL and OP data on the instability of $\ell=1$ and $\ell=2$ modes
of $\nu$\,Eri seismic model. The OPAL model was discussed by
Pamyatnykh et al.\,(2004). Using the OP data leads to wider
frequency instability range, as it has been demonstrated very
clearly for whole $\beta$\,Cep and SPB domains by Miglio et
al.\,(2007). 
They showed also that the effect of the choice between old and new
solar proportions in the heavy element abundances - respectively,
GN93 (Grevesse \& Noels 1993) and A04 (Asplund et al.\,2005) - is
significantly smaller than that of the opacity choice. Only the
model constructed with the OP data is unstable in low-frequency
modes (high-order g modes of $\ell=2$) and can fit the measured
frequency at 0.61~c/d. The local $\eta$ maximum for $\ell=1$ nearly
coincides with another measured low frequency at 0.43~c/d, but no
instability was found there. There is also a problem of excitation
and fitting of the observed peaks at 7.89~c/d, one of them was
identified from photometry as an $\ell=1$ mode (it must be $\ell=1,
{\rm p}_2$ mode). Zdravkov \& Pamyatnykh (these Proceedings)
demonstrated that an additional opacity enhancement by approximately
50\,\% around the iron bump at its slightly deeper location may lead to 
the instability of both high-order g modes and $\ell=1, {\rm p}_2$ mode
at the observed frequencies.

The two $\ell=1$ triplets in the $\nu$\,Eri frequency spectrum allow
to constrain internal rotation rate. 
In Fig.\,4, the rotational splitting kernels (weighting
functions in the integral over the local rotational velocity inside
the star) for two dipole modes of the seismic $\nu$\,Eri model are
shown. These kernels determine the sensitivity of different layers
to the rotational splitting of the oscillation frequencies,
\figureDSSN{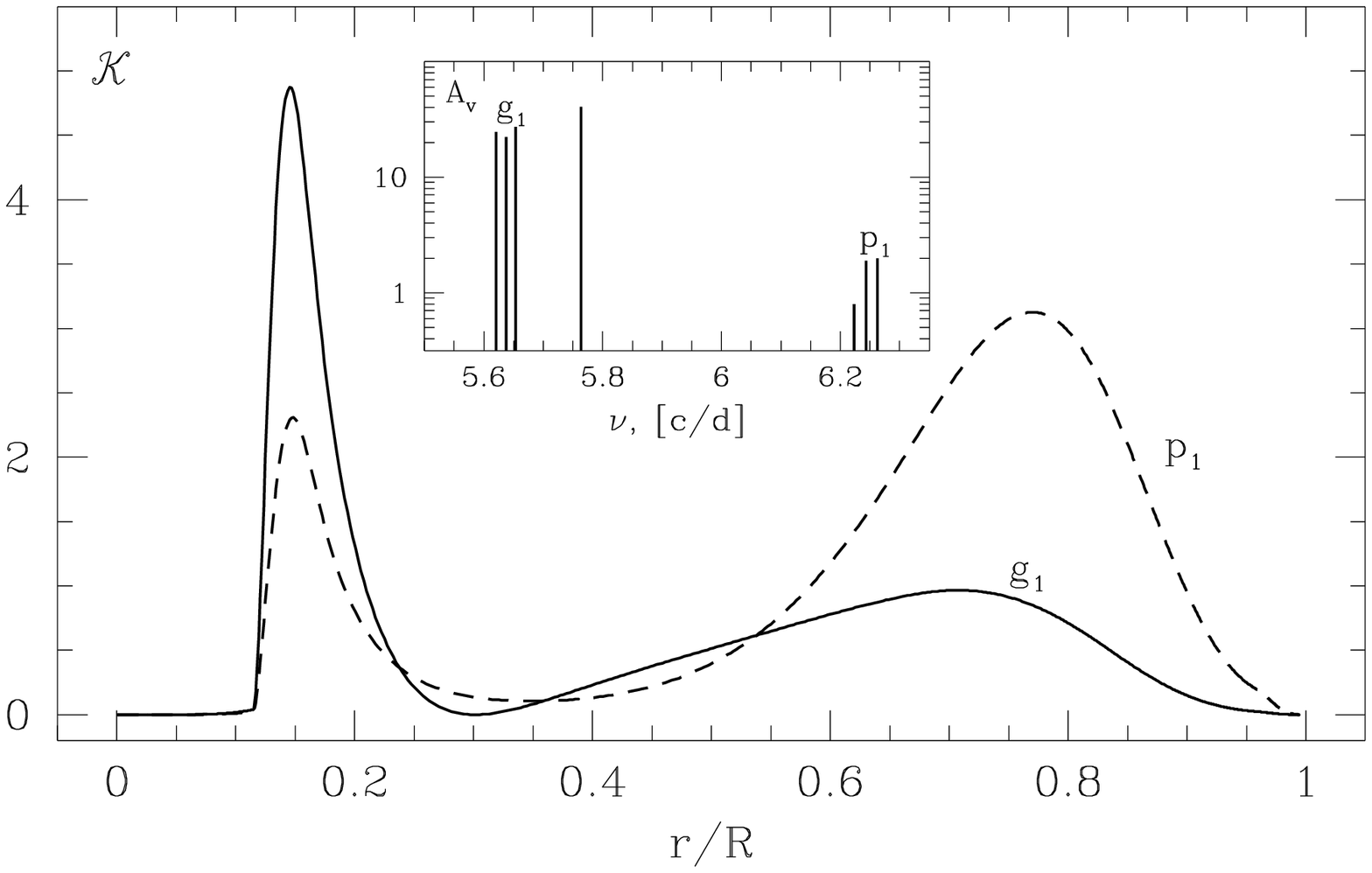}
{
The rotational splitting kernels, ${\cal K}$, for the
two dipole ($\ell=1$) modes in the seismic model of $\nu$\,Eri. The
corresponding part of the frequency spectrum is shown in the small
box. The inner maxima extend over the chemically
nonuniform zone above the convective core.
}
{RotKer_NuEri}{!ht}{clip,angle=0,width=75mm}
therefore from the measured spacings between components
of different multiplets we can obtain information about internal
rotation. We showed that in $\nu$\,Eri the convective core rotates
approximately 5 times faster than the envelope.


In Fig.\,5, the observed frequency spectrum of {\bf 12\,Lac} is
compared with results for a seismic model which was constructed to
fit four dominant peaks at 5.0-5.5~c/d: the radial fundamental mode,
two components of $\ell=1$ triplet and one component of the $\ell=2$
quintuplet. Most of remaining peaks also have their counterparts
among theoretical frequencies. The $\eta(\nu)$ dependence is similar
to that for the $\nu$\,Eri model in Fig.\,3. The theoretical
frequency range of unstable $\beta$\,Cep type modes fits the
observed peaks very well. However, the high-order g modes of
$\ell=1$ and $\ell=2$ are stable, in contrast to the $\ell=2$ case
of $\nu$\,Eri. Again, the solution of the problem may require an
opacity enhancement in the driving zones in deep envelope.
\figureDSSN{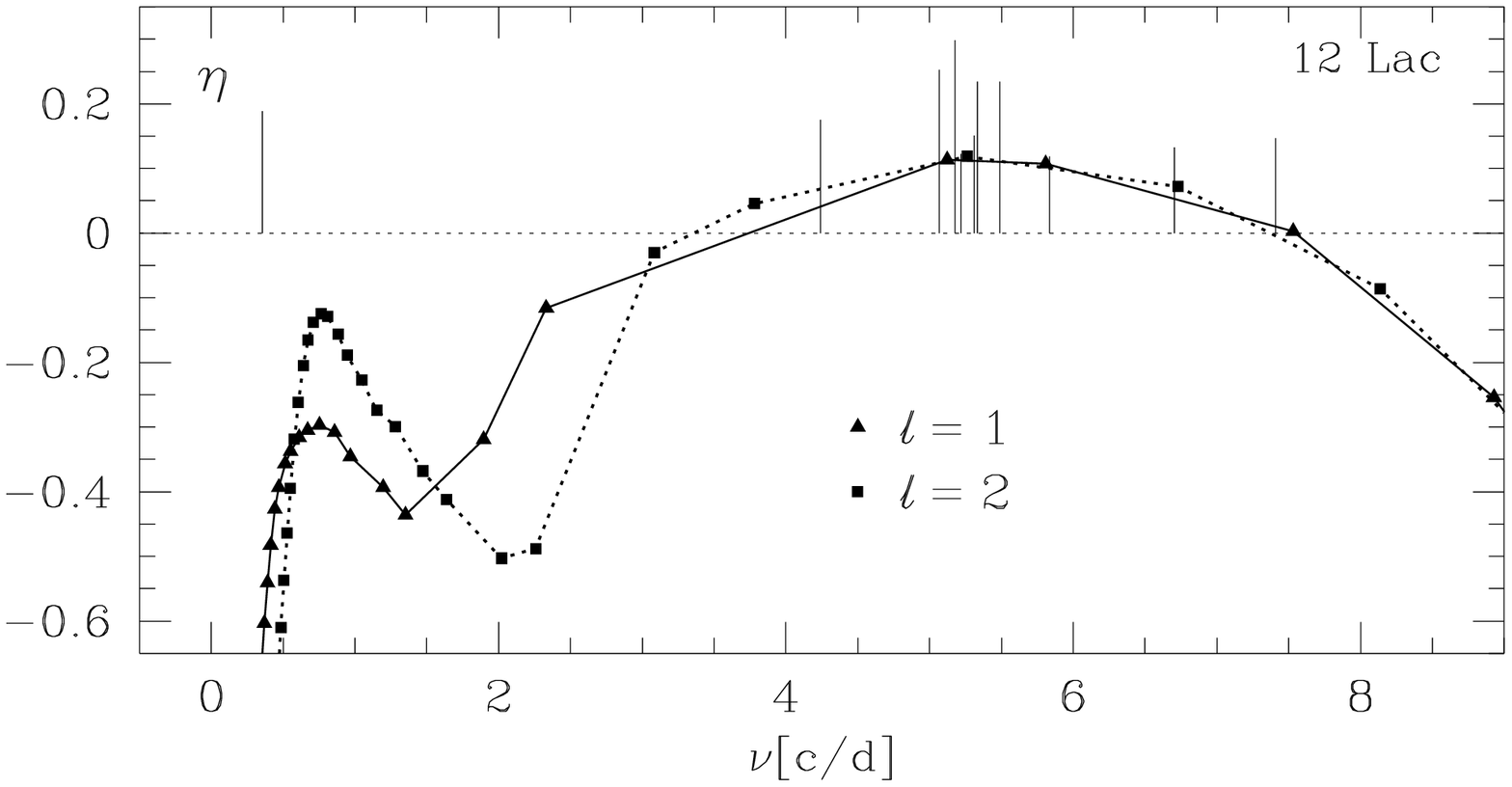}
{
The normalized growth rates, $\eta$,
of $\ell=1$ and $\ell=2$ modes as a function of mode frequency in
seismic model of {\bf 12\,Lac} computed with the OP opacities for $Z=0.015$
and heavy element mixture A04. Vertical lines mark the observed frequencies,
with amplitudes given in a logarithmic scale.
}
{Eta_12Lac}{!ht}{clip,angle=0,width=75mm}
\figureDSSN{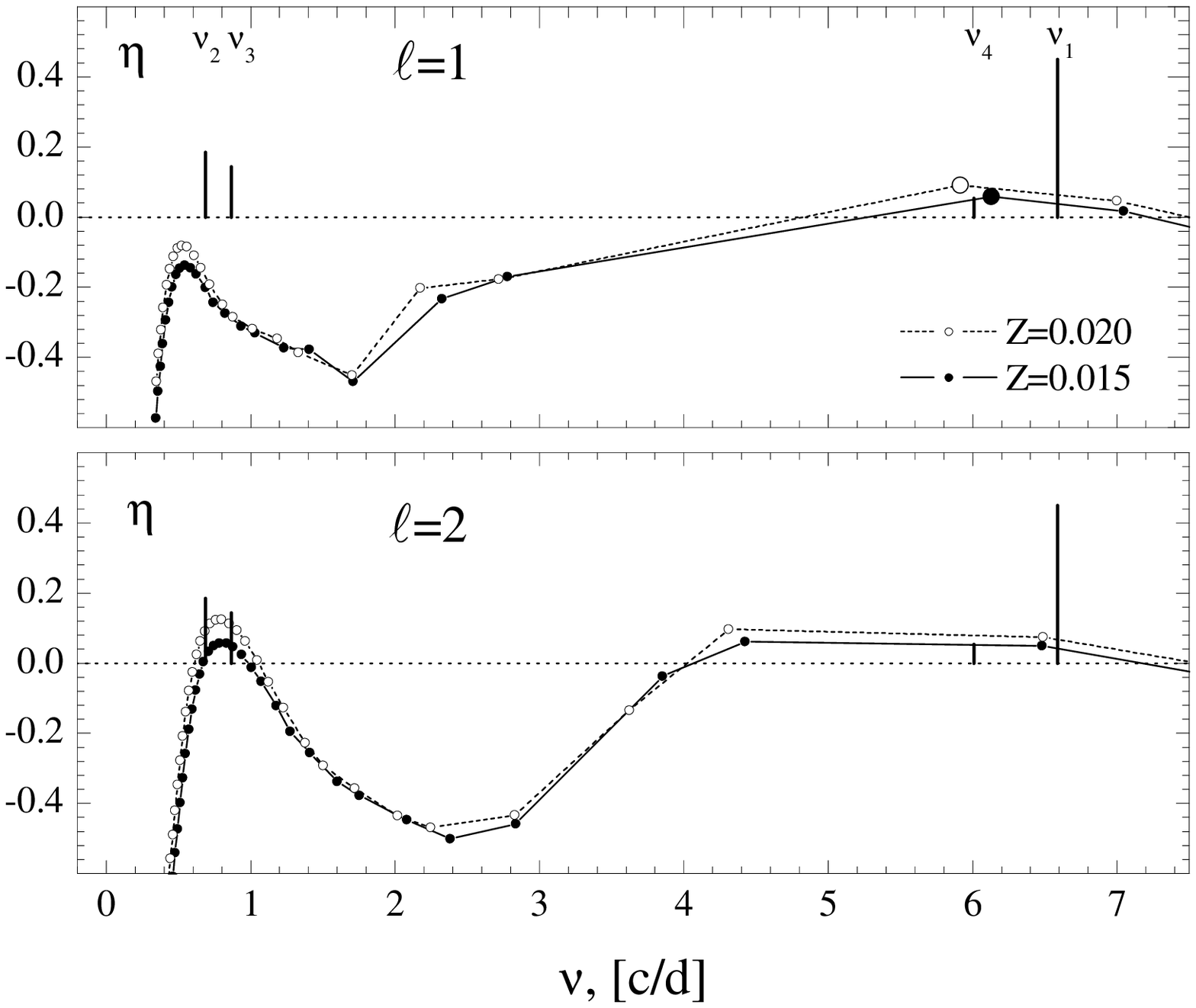}
{
The normalized growth rates of $\ell=1$ and 2 modes in two
{\bf $\gamma$\,Peg} models constructed with the OP opacities for the
heavy element mixture A04. Both models have mass of 8.7~$M_{\odot}$
but differ in the heavy element mass fraction, $Z$. Two larger points
close to the $\nu_4$ mark mode $\ell=1, {\rm g}_1$.
}
{Eta_GamPeg}{!h}{clip,angle=0,width=75mm}
From the frequencies of two components of rotationally splitted $\ell=1$ triplet (and taking
into account a necessity to fit the observed $\ell=2$ peak at 4.2 c/d) we estimated
that the convective core rotates approximately 4.6 times faster than the envelope,
this result is similar to that for $\nu$\,Eri. The calculated surface equatorial
velocity of 47\,km/s agrees very well with the spectroscopic estimation of about 50\,km/s.


{\bf $\gamma$\,Peg} is also a hybrid star (Chapellier et al.\,2006
and private communication by G.\,Handler) with two oscillation
frequencies of the $\beta$\,Cep type (6.01 and 6.59~c/d), and two -
of the SPB type (0.686 and 0.866~c/d). Fig.\,6 illustrates
preliminary results of the modelling. The models were constructed by
fitting the frequency of the radial fundamental mode to the observed
frequency at 6.59~c/d. As it is easy to see from Fig.\,6, the
frequency of dipole mode $\ell=1, {\rm g}_1$ can fit the observed
value 6.01~c/d in a model with $Z$ value in between 0.015 and 0.02.
Two observed low-frequency modes are well inside the frequency range
of unstable high-order gravity modes of $\ell=2$. The identification
of the mode degree for observed peaks will allow to check these
preliminary assessments. Now, an intensive campaign of ground-based 
and cosmic observations (with the MOST satellite) is on the way 
(G.\,Handler, private communication).

\section*{44\,Tau}
The $\delta$\,Sct type variable 44 Tau is an extremely slowly
rotating star in which 15 oscillation frequencies have been measured
(Antoci et al.\,2007, Breger \& Lenz 2008).
\figureDSSN{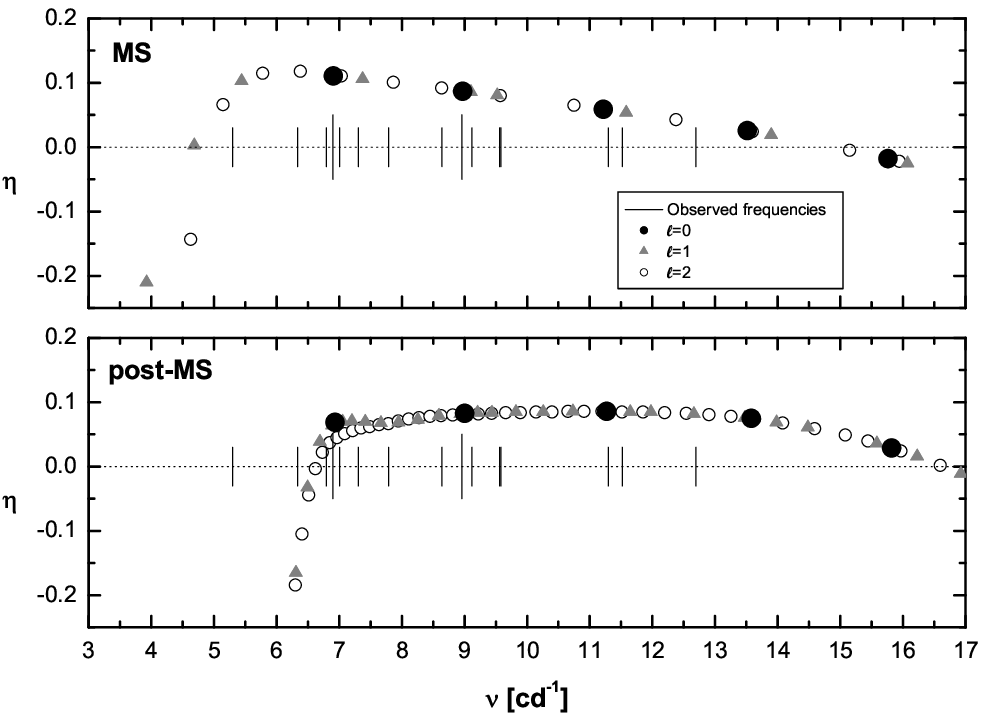}
{
The normalized growth rates of $\ell=0-2$ modes in
the main-sequence and post-MS models of {\bf 44\,Tau}.
Vertical lines mark the observed frequencies, with longer
lines for two identified radial modes. Both models fit these
radial frequencies.
}
{Eta_44Tau}{!h}{clip,angle=0,width=75mm}
Two of them are identified as radial modes. Moreover, 7 other modes
have been identified as $\ell=1$ and $\ell=2$ oscillations. The
$\log g$ value as determined from photometry and spectroscopy does
not allow to choose between the main-sequence and post-MS model for
this star. Lenz et al.\,(2008) gave asteroseismic arguments in favor
of the post-MS model. In particular, only post-MS model can fit both
observational error box in the HR diagram and frequencies of two
radial modes. In Fig.\,7 the observed and theoretical frequency
ranges are compared for two 44 Tau star models. In contrast to a
similar plot given by Lenz et al.\,(2008), this figure also includes
the previously unknown lowest frequency mode at 5.30~c/d which has
been detected very recently (Breger \& Lenz 2008). This finding may
be an argument against the post-MS models for
44\,Tau which have no unstable modes at these low frequencies.

\acknowledgments
{
A partial financial support from the Polish MNSiW
grant No. 1 P03D 021 28 and from the HELAS project is acknowledged.
I am grateful to Tomasz Zdravkov for Figs.\,1 and 6,
and to Patrick Lenz for Fig.\,7 and useful comments.
}

\References
{
\rfr Antoci~V., Breger~M., Rodler~F., et~al. 2007, A\&A, 463, 225.
\rfr Ausseloos~M., Scuflaire~R., Thoul~A., Aerts~C. 2004, MNRAS, 355, 352.
\rfr Asplund~N., Grevesse~N., Sauval~A.\,J. 2005, ASP Conf.\,Ser., vol.\,336, p.\,25.
\rfr Breger~M., \& Lenz~P. 2008, A\&A, 488, 643.
\rfr Chapellier~E., Le~Contel~D., Le~Contel~J.\,M., et~al. 2006, A\&A, 448, 697.
\rfr De~Cat~P., Briquet~M., Aerts~C., et~al. 2007, A\&A, 463, 243.
\rfr Dziembowski~W.\,A., \& Pamyatnykh~A.\,A. 1993, MNRAS, 262, 204.
\rfr Dziembowski~W.\,A., \& Pamyatnykh~A.\,A. 2008, MNRAS, 385, 2061.
\rfr Grevesse~N., \& Noels~A. 1993, in {\it Origin and Evolution of
  the Elements}, eds.\,Pratzo~N., Vangioni-Flam~E., Casse~M.,
  Cambridge Univ.\,Press., p.\,15.
\rfr Handler~G., Jerzykiewicz~M., Rodriguez~E., et~al. 2006, MNRAS, 365, 327.
\rfr Iglesias~C.\,A., \& Rogers~F.\,J. 1996, ApJ, 464, 943.
\rfr Jerzykiewicz~M., Handler~G., Shobbrook~R.\,R., et~al. 2005, MNRAS, 360, 619.
\rfr Lenz~P., Pamyatnykh~A.\,A., Breger~M., \& Antoci~V. 2008, A\&A, 478, 855.
\rfr Miglio~A., Montalb\'an~J., \& Dupret~M.-A. 2007, MNRAS, 375, L21.
\rfr Pamyatnykh~A.\,A. 1999, Acta Astr., 49, 119.
\rfr Pamyatnykh~A.\,A., Handler~G., \& Dziembowski~W.\,A. 2004, MNRAS, 350, 1022.
\rfr Seaton~M.\,J. 2005, MNRAS, 362, L1.
\rfr Stankov~A., \& Handler~G. 2005, ApJSuppl., 158, 193.
}

\begin{center}\textbf{DISCUSSION}\end{center}
\noindent \discussion
{Arlette Noels} 
{I fully agree with you
that there probably is a missing opacity not only in $\beta$ Cep stars
but all along the MS.  I am however surprised by your results for $\nu$
Eri on the comparison between OP and OPAL opacities. The "best"
models you obtain are very different which means that the change is
not only due to changes in the iron bump but also to changes in the
whole star. Do you agree with that? This is very different from an
increase of opacity limited to the iron bump which will only affect
the excitation of the modes.}
\discussion{Alexey Pamyatnykh} 
{You are correct in that our "best" seismic models computed with 
the OPAL and OP data differ in a some extent not only in the iron bump region 
but also in the whole star. The most important difference between 
the OP and OPAL opacities is that in the OP case the iron bump is located 
slightly deeper in the star (at slightly higher temperatures) compared 
with the OPAL case. Our seismic models of $\nu$\,Eri fit the observed 
frequencies of the radial fundamental mode and two dipole modes 
(${\rm g}_1$ and ${\rm p}_1$) very nicely, with an accuracy better than $10^{-3}$. 
The opacity changes around the iron bump even by about 20 percent 
(which is a systematic difference between OP and OPAL at fixed temperature 
in this region) lead to a change of the stellar structure and consequently 
to a change of the oscillation frequency spectrum, so to fit the observed 
frequencies we must adjust global stellar parameters - such as mass, 
heavy element abundance and effective temperature. 
Therefore, the structure of the seismic model constructed with the OP data will 
differ in a some extent from that of the model constructed with  OPAL.  

\noindent Again, an additional similar opacity change around the iron bump will also affect 
the oscillation frequencies (not only the mode excitation) and thereby 
the corresponding seismic model. Such a test with artificially enhanced opacity 
in a vicinity of the iron bump (more exactly, slightly deeper than the iron bump!) 
has been performed by T. Zdravkov and myself (these Proceedings).}

\end{document}